\documentclass[aps,twocolumn,floats,nofootinbib,prl]{revtex4}
\usepackage{graphics,graphicx,epsfig}
\usepackage{amssymb,color}
\usepackage{epsf,epstopdf,wrapfig}
\usepackage[intlimits]{amsmath}
\usepackage{textgreek}
\usepackage{tikz-cd}
\usepackage{lipsum}

\newcommand{\beq}{\begin{equation}}
\newcommand{\eeq}{\end{equation}}
\newcommand{\beqn}{\begin{eqnarray}}
\newcommand{\eeqn}{\end{eqnarray}}

\usepackage{mathtext}
\usepackage[active]{srcltx}
\usepackage[T2A]{fontenc}
\usepackage[utf8]{inputenc}
\usepackage[english]{babel}
\usepackage{amsfonts}
\usepackage[intlimits]{amsmath}
\usepackage{amsthm}
\usepackage{amssymb}
\usepackage{amscd}
\usepackage{enumerate}
\usepackage{bbm}
\usepackage{dsfont}
\usepackage{slashed}
\usepackage[normalem]{ulem}
\usepackage{ulem}
\usepackage[dvipdf]{feynmf}
\usepackage{verbatim}
\usepackage{prettyref}
\usepackage{graphicx}
\usepackage{setspace}
\usepackage{esint}
\usepackage{marvosym}

\allowdisplaybreaks[1]

\usepackage{indentfirst}

\usepackage{hyperref}
\hypersetup{
    pdfauthor={Vasyl Alba},
    colorlinks = false,
    linkcolor = black,
    anchorcolor = red,
    citecolor = black,
    filecolor = red,
    urlcolor = blue,
    unicode = true
}

\begin{document}

\title{Exploring a strongly non--Markovian animal behavior}

\author{Vasyl Alba,$^{1,2}$ Gordon J. Berman,$^{1,3}$ William Bialek,$^{1,4}$ and Joshua W. Shaevitz$^1$}

\affiliation{$^1$Joseph Henry Laboratories of Physics and Lewis--Sigler Institute for Integrative Genomics, Princeton University, Princeton NJ 08544 \\
$^2$Department of Engineering Science and Applied Mathematics and NSF--Simons Center for Quantitative Biology, Northwestern University, Evanston IL 60208\\
$^3$Departments of Biology and Physics, Emory University, Atlanta GA 30322 \\
$^4$Initiative for the Theoretical Sciences, The Graduate Center, City University of New York, 365 Fifth Avenue, New York NY 10016}

\begin{abstract}
A freely walking fly visits roughly 100 stereotyped states in a strongly non-Markovian sequence.   To explore these dynamics, we develop a generalization of the information bottleneck method, compressing the large number of behavioral states into a more compact description that maximally preserves the correlations between successive states.   Surprisingly, preserving these short time correlations with a compression into just two states captures the long ranged correlations seen in the raw data.  Having reduced the behavior to a binary sequence, we describe the distribution of these sequences by an Ising model with pairwise interactions, which is the maximum entropy model that matches the two-point correlations. Matching the correlation function at longer and longer times drives the resulting model toward the Ising model with inverse square interactions and near zero magnetic field.  The emergence of this  statistical physics problem from the analysis real data on animal behavior is unexpected.
\end{abstract}

\maketitle

In the twentieth century there were two very distinct approaches to the characterization of animal behavior.  Ethologists focused on behavior in its natural context, often describing in qualitative terms phenomena of great complexity \cite{frisch_74,gould_82}. Psychophysicists, in contrast, brought behavior into the laboratory, constraining subjects to choosing from a small set of alternatives, quantifying the probabilities of different choices as a function of sensory inputs and task constraints  \cite{lawson+uhlenbeck_50,green+swets_66}.  Recently  there has been considerable interest in quantifying unconstrained and more naturalistic behaviors.  Examples include exploring the variability of eye movement trajectories in primates \cite{osborne+al_05}, and the postural dynamics of freely moving {\em C elegans} \cite{stephens+al_08,ahamed+al_20},  walking flies \cite{branson+al_09,berman+al_14,berman+al_16}, and mice \cite{wiltschko+al_15,marshall+al_20}.  The combination of high resolution video imaging and efficient AI tools \cite{mathis+al_18,pereira+al_19} is making these approaches more generally applicable, and these data have focused attention on the wide range of time scales in the behavior of single organisms, from milliseconds to a  lifetime
\cite{brown+bivort_17,datta+al_19,pereira+al_20,mathis+mathis_20,bialek_20}.

The behavior of a fly walking freely in a featureless arena can be described as a sequence of transitions among discrete, stereotyped states \cite{berman+al_14}.  Some of these states correspond to  actions with a simple verbal description, such as grooming particular body parts, while other states are not so simple; nonetheless all flies of a single species revisit the same $\sim$100 states, and one can recognize the same states in closely related species.  Although the description of behavior as a sequence of states often is accompanied an analysis of transitions from one state to the next, and the (possibly implicit) hypothesis that these transitions are independent of one another, fly behavior dramatically violates this Markovian assumption \cite{berman+al_16}.      We would like to find a simple phenomenological description of behavioral sequences that captures this non--Markovian structure, ideally leading to some insight into the internal states that make it possible.

The non--Markovian character of the fly's trajectory through state space can be seen in several ways.  In particular, if we look at the probability that the fly moves from one state to another after a time $\tau$, then in a Markov model all of these transition matrices are powers of the fundamental transition matrix describing the model, and the eigenvalues of this matrix will decay at a constant rate.  This predicts that essentially all memory of the initial state will be lost after $\sim 30$ transitions.  Instead, memory persists out to $\sim 1000$ transitions, and the eigenvalues of the transition matrix decay more and more slowly as we look across longer time scales \cite{berman+al_16}.

\begin{figure}[t]
\centerline{\includegraphics[width = \linewidth]{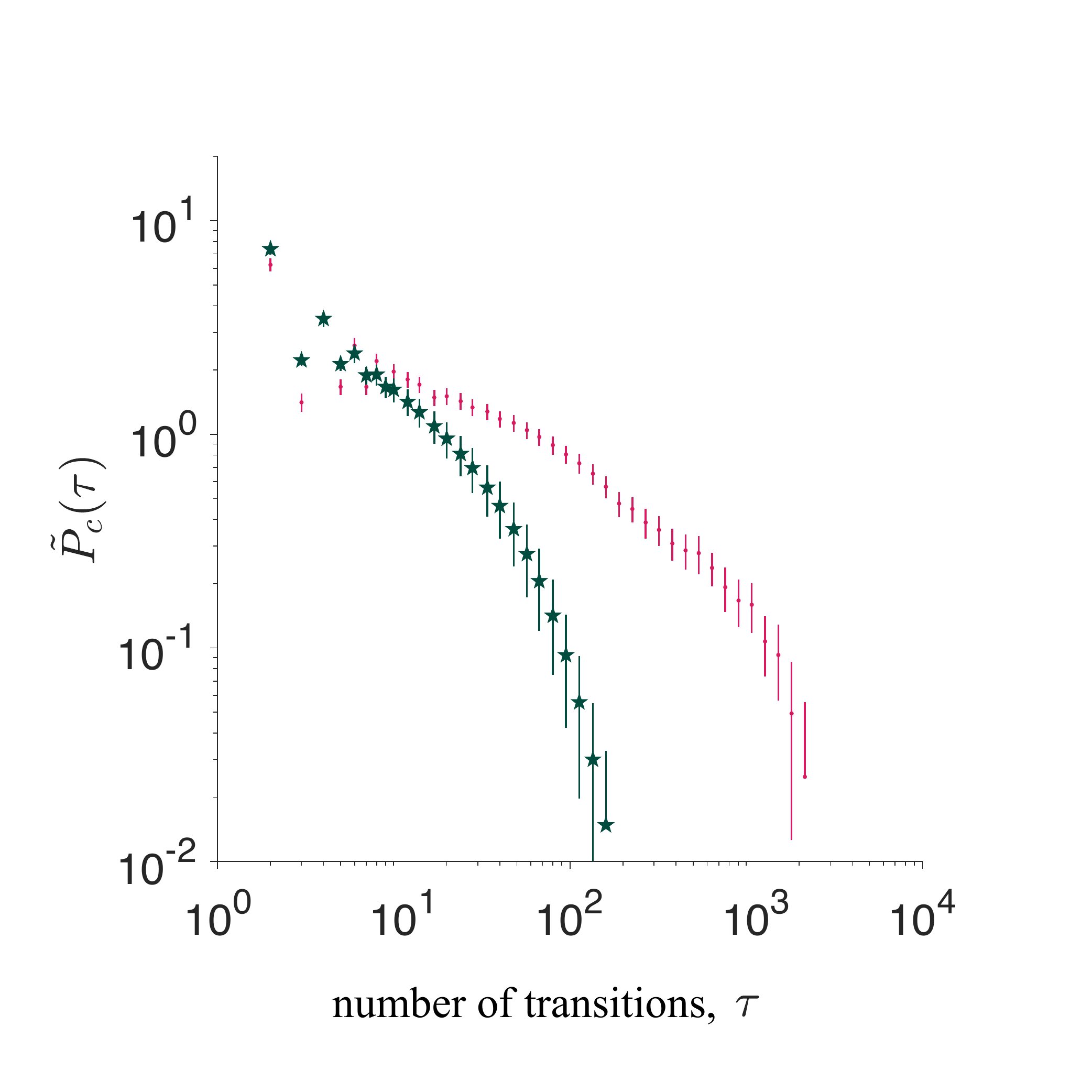}}
\vskip  -0.25 in
\caption{Probability $P_c(\tau)$ that a behavioral state is revisited after $\tau$ transitions. We normalize so that ${\tilde P}_c(\tau ) = P_c(\tau )/P_c(\tau\rightarrow\infty ) - 1$ decays to zero.  Means and standard errors across 59 flies (red) compared with predictions from a Markov model that matches observed transition probabilities (green).  In both cases we analyze individual flies and average ${\tilde P}_c$ at the end. 
\label{Pc}}
\end{figure}

We can see the same effect in the mutual information between states separated by a time $\tau$, or in  the probability that the fly returns to the same state after $\tau$ transitions, $P_c(\tau )$, shown in Fig \ref{Pc}.  In both of these representations we  detect correlations in behavioral state extending over thousands of transitions, far beyond what is predicted by a Markov model matched to the measured transition probabilities at $\tau =1$.  More subtly, the decay of mutual information or return probability is gradual, and does not seem to be characterized by discrete time scales, echoing the  slowing of eigenvalue decays.

The problem that we face in characterizing the non--Markovian structure of behavior is that with more than one hundred states, there are $\sim 10^4$ possible transitions between pairs of states.  If we observe a single fly for one hour, we see roughly this number of transitions \cite{berman+al_14,berman+al_16}.  If we  make much longer observations on single flies we encounter  non--stationarity, while if we merge data from many flies  we may obscure individual differences.  To make progress we need to simplify our description of the behavioral states, but we need to do this in a way that preserves the long memory seen in the full description.

\begin{figure}[b]
\centerline{\includegraphics[width = 0.75\linewidth]{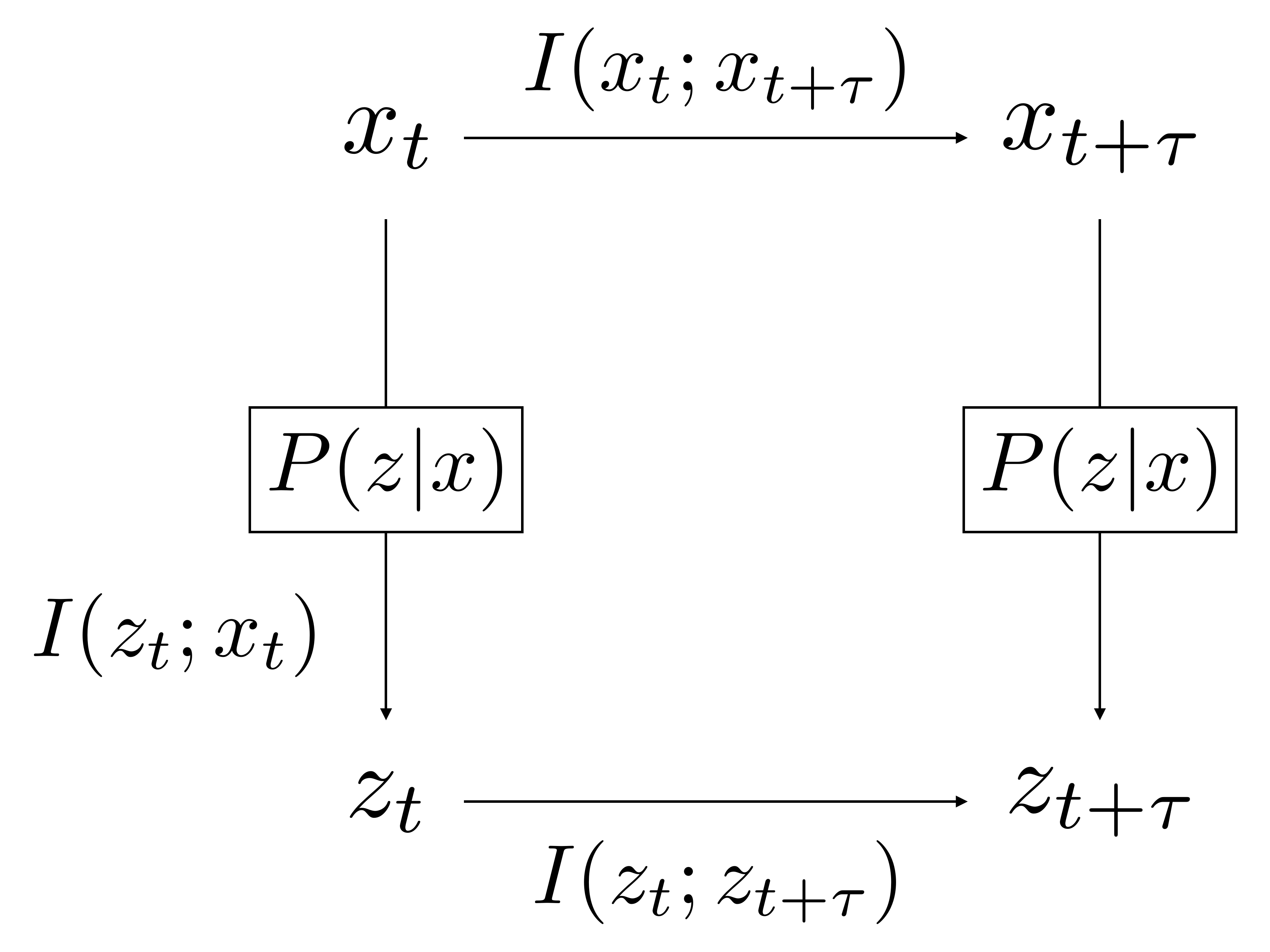}}
\caption{Mapping of behavioral states $x$ into a compressed representation $z$.  We compress the state at each moment in time independently, with a mapping $x\rightarrow z$ that is in general probabilistic, $P(z|x)$.  The compressed representation captures an information $I(z_t; x_t)$ about the original state, and the compressed variables share information $I(z_t; z_{t+\tau})$ across time $\tau$.  We choose the compression to maximize $I(z_t; z_{t+\tau})$ at fixed $I(z_t; x_t)$, as in Eq (\ref{eq1}). \label{fig:diagram}}
\end{figure}

Concretely, we want to map the behavioral state at each moment in time into some reduced or clustered description, $x_t \rightarrow z_t$, with the hope that we can preserve the long memories that we see in the state sequence; in general this mapping can be probabilistic, described by $P(z_t | x_t)$. The ``information bottleneck'' problem \cite{tishby+al_99} is the search for a compression that maintains information about some other variable, which we could take to be a behavioral state in the future \cite{berman+al_16}, but here we want a more symmetric formulation (Fig \ref{fig:diagram}).

We  choose compressions  that preserve temporal correlations  by maximizing the mutual information between compressed variables at different times, $I(z_t ; z_{t+\tau})$.  To control the complexity of our description, we limit the information that we capture about the original variables, $I(z_t ; x_t)$.  In principle the mapping $x_t \rightarrow z_t$ could be probabilistic, so we  solve the variational problem
\begin{equation}
\max_{P(z | x)} \left[ I(z_t ; z_{t+\tau}) - T I(z_t ; x_t) \right] ,
\label{eq1}
\end{equation}
where the Lagrange multiplier $T$ imposes the constraint on $I(z_t ; x_t)$.  As in the original bottleneck problem, $T$ plays the role of a temperature, such that as $T \rightarrow 0$ the mapping $x\rightarrow z$ becomes deterministic.

Following the same steps as in Ref \cite{tishby+al_99}, we find that the solution to the maximization problem in Eq (\ref{eq1}) obeys the self--consistent equation
\begin{eqnarray}
P(z|x) &=& \frac{P(z)}{{\cal Z}(x;T)}\exp \left[ -\frac{1}{T} F(x, z)\right],\\
F(x_t, z_t) &=&  D_{KL}\big[ P(z_{t+\tau}|x_t) || P(z_{t+\tau}|z_{t})\big] \nonumber\\
&&\,\,\,\,\,\,\,\,\,\, +D_{KL}\big[ P(z_{t+\tau}|x_t) || P(z_{t+\tau}|z_t)\big],
\end{eqnarray}
where $D_{KL}[P || Q]$ is the Kullback--Leibler divergence between the distributions $P$ and $Q$, and ${\cal Z} (x;T)$ is a normalization constant.  Again following Ref \cite{tishby+al_99}, we can turn a self--consistent equation into an iterative algorithm.  We have explored solutions with different cardinality for $z_t$, and find that they form a hierarchical clustering scheme, as expected from earlier work \cite{berman+al_16}.  Our focus here is on the most severe compression, in which $z_t$ has just two states, and the limit $T\rightarrow 0$, where the mapping $x\rightarrow z$ becomes deterministic.  Further, we search for compressions that preserve information from one state to the next, corresponding to $\tau = 1$ in Eq (\ref{eq1}).

\begin{figure}[t]
\centerline{\includegraphics[width = \linewidth]{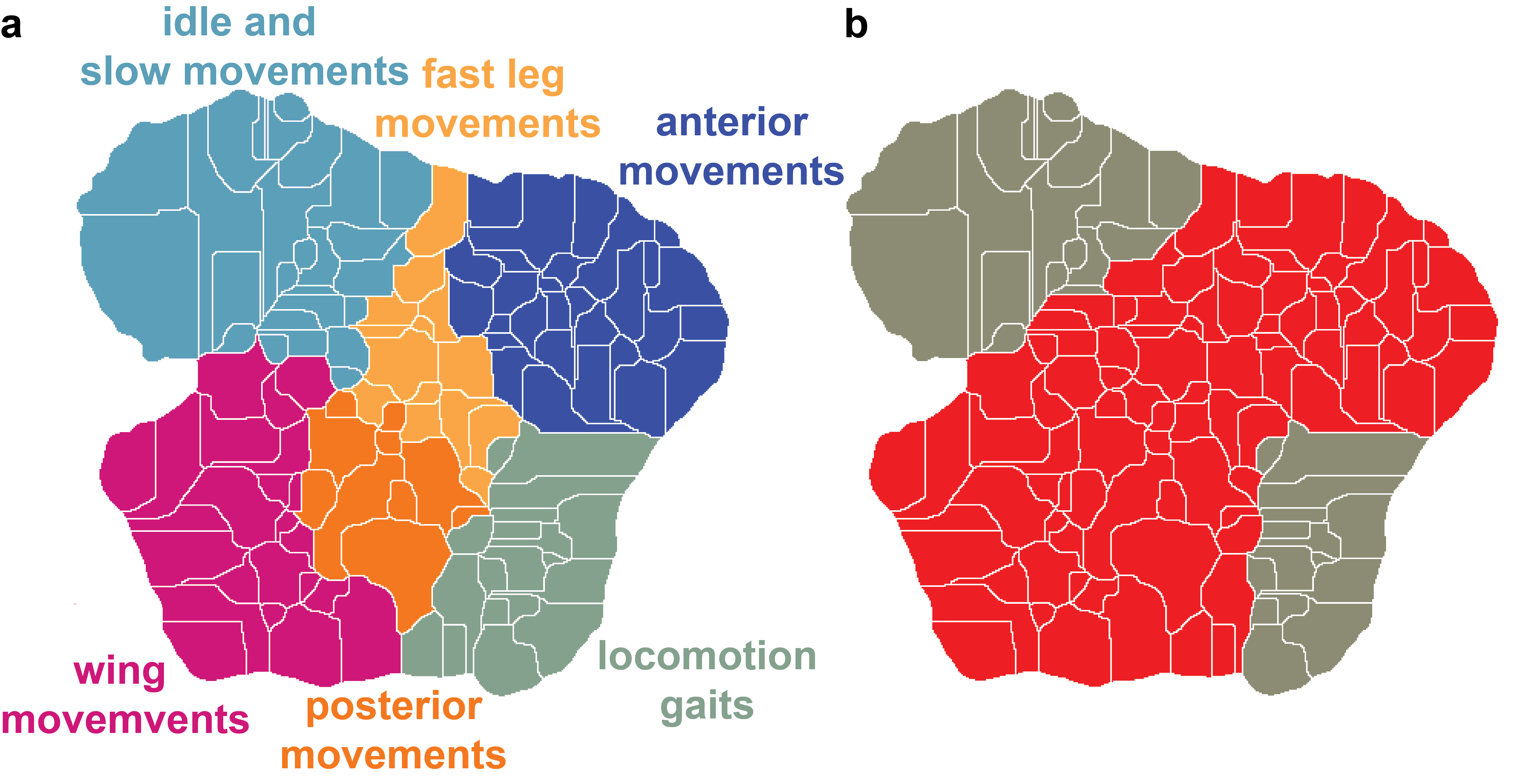}}
\caption{Behavioral states and their compressed representation.  {\bf (a)} The original $N_s = 123$ states and their boundaries, grouped informally into clusters \cite{berman+al_14}. {\bf (b)}  The optimal compression into two states, with $\sigma = +1$ shown in red and $\sigma = -1$ shown in grey.
\label{states}}
\end{figure}

When we compress into just two states we turn behavior into a binary sequence, or equivalently a one--dimensional chain of Ising spins $z_t = \sigma_t = \pm 1$;  the structure of this mapping is illustrated in Fig \ref{states}.   Each point in the two--dimensional behavioral space represents a short temporal sequence from the original video recording; the probability distribution in this plane has many well resolved peaks, and the dynamics consist of sojourns in the neighborhood of one peak followed by a quick jump to another, allowing  us to define the $N_s = 123$ discrete states and their boundaries \cite{berman+al_14}.  States which are neighbors in these two dimensions are similar by construction, and can be (informally) grouped into the clusters shown in Fig 3a--anterior movements, posterior movements, locomotion gaits, etc.   In Fig 3b we have the results of the optimal mapping $x\rightarrow z$, which groups together a range of relatively rapid movements into one state $\sigma = +1$ while walking and idling are mapped to $\sigma = -1$  \cite{difference}.

\begin{figure}[b]
\vskip  -0.25 in
\centerline{\includegraphics[width = \linewidth]{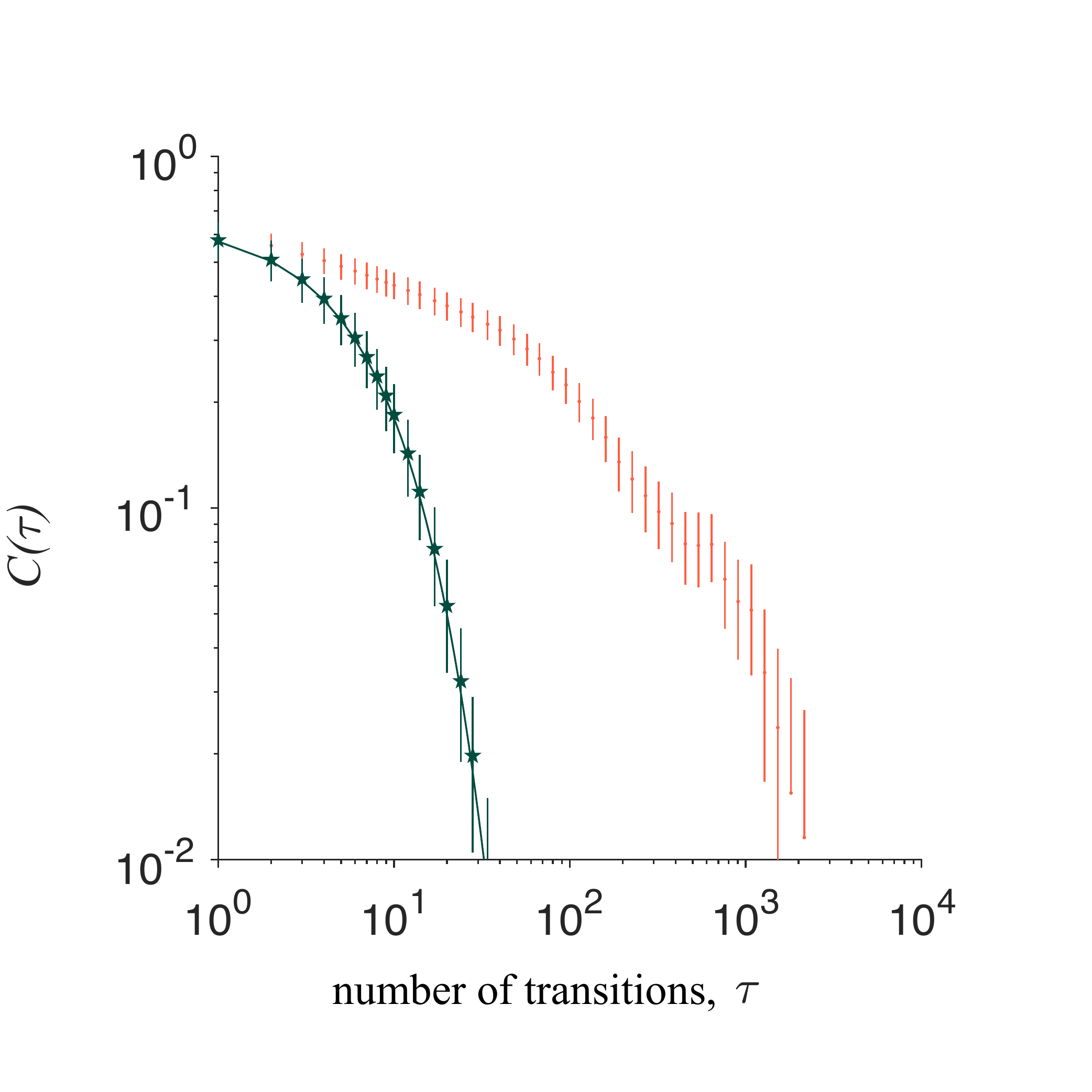}}
\vskip  -0.25 in
\caption{Connected correlation function for the binary description of behavior, Eq (\ref{Ctau}). Data points and error bars (red) as in Fig 1, compared with a Markov model over the two states (green).  Solid line is a single exponential decay with correlation time $\tau_c = 7.9\pm 0.9$.  \label{ss}}
\end{figure}

Once we have reduced behavioral trajectories to a chain of Ising spins, we can measure temporal correlations with the usual spin--spin correlation function \cite{Pc_C},
\begin{equation}
C(\tau) = \langle \sigma_t\sigma_{t+\tau}\rangle - \langle \sigma_t\rangle^2 .
\label{Ctau}
\end{equation}
Perhaps surprisingly, although we  asked for a compression  that preserves information only across one behavioral transition, Fig \ref{ss} shows that the resulting binary or Ising variables have measurable correlations out to $\sim$1000 transitions.    While the compression into two behavioral ``states'' obviously throws away a lot of detail (by construction), long time correlations are preserved.

Having reduced behavior to binary sequences, we would like to understand  the full probability distribution for these sequences, $P(\{\sigma_t\})$, going beyond the Markov approximation. We want our approximation to match the observed bias between the two behavioral states, measured by the expectation value $\langle \sigma_t \rangle$.  We also want to capture the long--ranged correlations in Fig \ref{ss}, so we insist that the correlation function $C(\tau)$ computed from the model $P(\{\sigma_t\})$ match the correlation function that we observe experimentally.  The minimally structured, or maximum entropy   model that matches $\langle \sigma_t \rangle$ and $C(\tau )$ is
\begin{equation}
P(\{\sigma_t\}) = {\frac{1}{Z}} \exp\left[ h\sum_{t} \sigma_t + {\frac{1}{2}} \sum_{t,t'} \sigma_t J(t-t') \sigma_{t'} \right] ,
\label{Pmaxent}
\end{equation}
where the interactions $J(\tau)$ must be tuned to match the correlation function $C(\tau)$, and the field $h$ must be tuned to match the asymmetry $\langle\sigma_t\rangle$ \cite{jaynes_57,bialek_12}.  We solve this inverse problem following the same strategy as in previous work \cite{tkacik+al_14}, using Monte Carlo to estimate  $C(\tau )$ in the model and adjusting $J(\tau )$ in proportion to the difference between this estimate and the measured values.  

\begin{figure}[t]
\vskip  -0.25 in
\includegraphics[width = \linewidth]{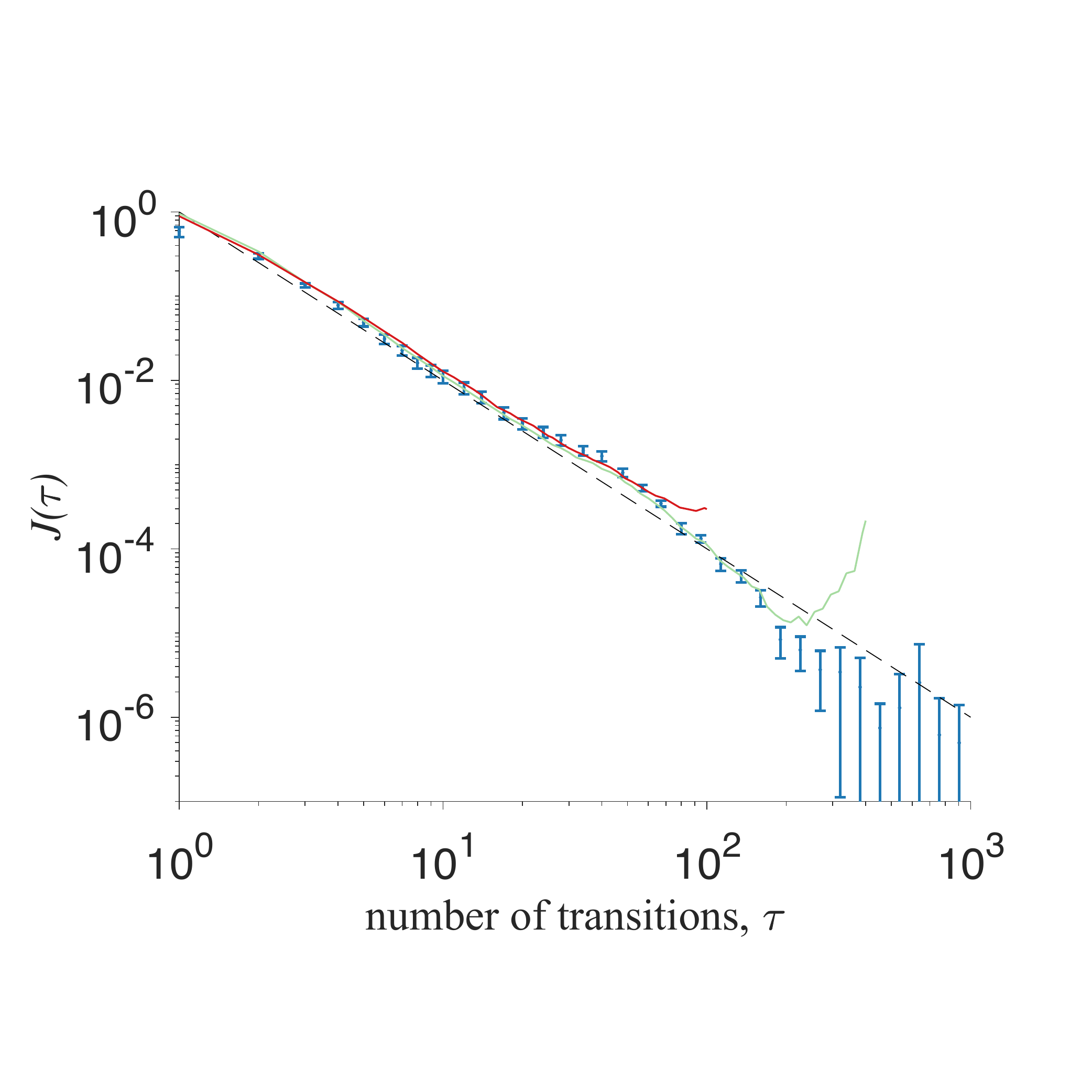}
\vskip  -0.5 in
\caption{Interactions  $J(\tau)$ needed to reproduce the correlation function $C(\tau )$ for $\tau < \tau_{\rm max}$, with $\tau_{\rm max} = 100$ (red), $400$ (green), $1000$ (blue points with errors).  Error bars are the standard deviations  across models that match $C(\tau)$ from randomly chosen halves of the flies in the experimental population.  Black dashed line is $J(\tau) = 1/\tau^2$, for comparison.  \label{J}}
\end{figure}

Even without detailed calculation, we know that generating long--ranged correlations in one dimension (here, time) requires long--ranged interactions, and this is what we find (Fig \ref{J}).   Specifically, if we try to match the correlations $C(\tau)$ for $|\tau| \leq \tau_{\rm max}$, then $J(|\tau| > \tau_{\rm max}) = 0$, but as we increase $\tau_{\rm max}$ we ``uncover''  interactions that couple states with longer and longer separations.  We see no end to this, within the limits of our data.   More subtly, although the asymmetry in the two clusters of behavioral states is large, corresponding to a ``magnetization'' $\langle \sigma_t \rangle = 0.424 \pm 0.048$, the magnetic field that we find is very small, $h = (4.7  \pm  0.7)\times 10^{-3}$ when we match correlations out to $\tau_{\rm max} = 10^3$.  Intuitively, long--ranged correlations imply that sequences of states are moving collectively, and hence  a small intrinsic bias is amplified.  

Over more than two decades in $\tau$, the interactions are very nearly $J(\tau ) =J_0 /\tau^2$.  The Ising model with such ``inverse square'' interactions has a fascinating history, and played an important role in the development of scaling ideas that presaged the full development of the renormalization group \cite{anderson_84}. It is quite startling to see this model emerge from the analysis of data on animal behavior.

\begin{figure}[b]
\vskip  -0.1 in
\centerline{\includegraphics[width = \linewidth]{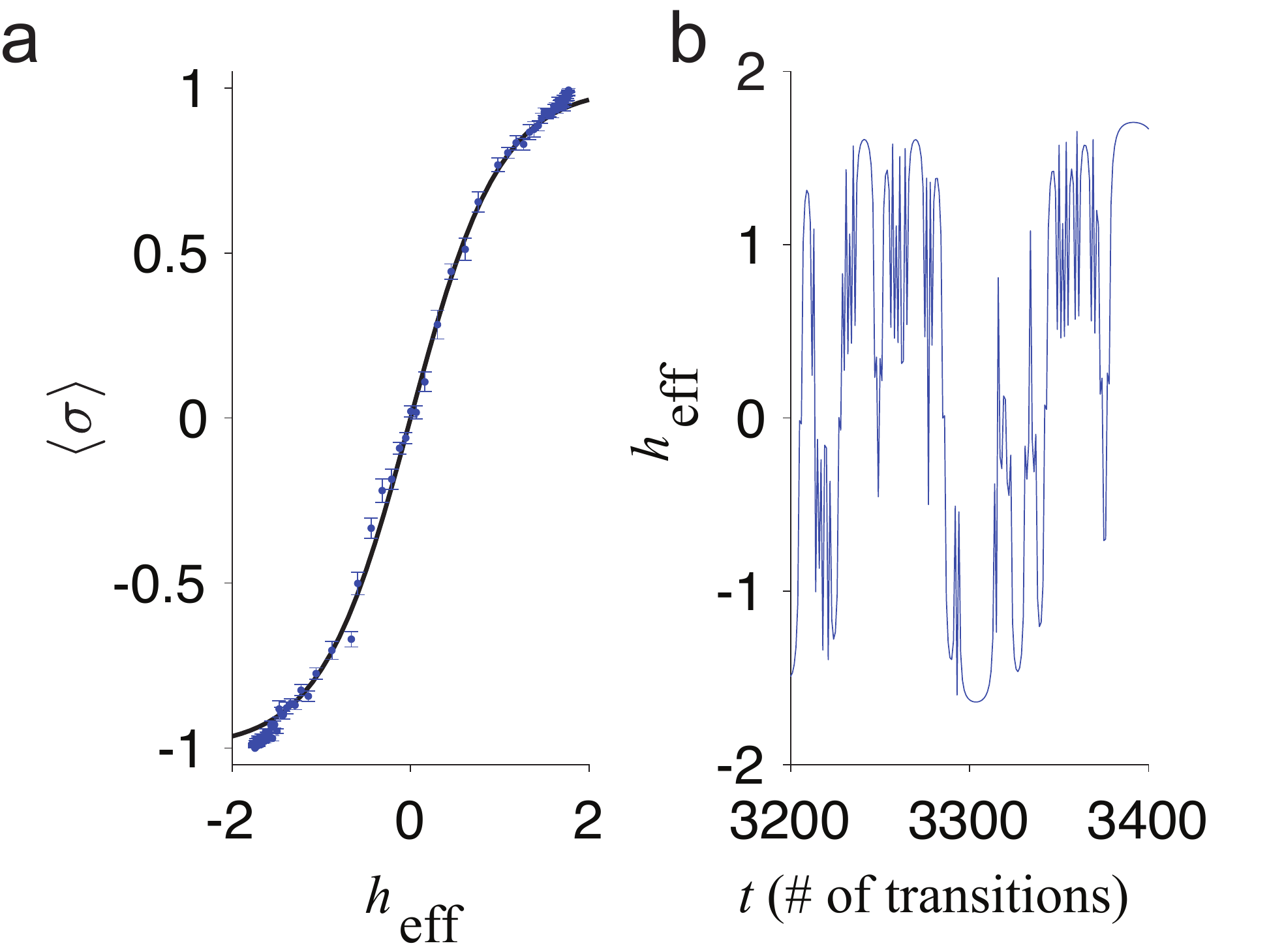}}
\caption{{\bf (a)} Mean behavioral state at one moment in time vs the effective field determined by states at other times, from Eq (\ref{heff}) with  $J(\tau ) = J_0/\tau^2$, $J_0 = 0.54\pm 0.01$.  Points are means across 59 flies, and error bars are standard deviations across random halves of this population; line is $\langle \sigma \rangle = \tanh(h_{\rm eff})$. {\bf (b)} A short segment of  $h_{\rm eff}$ vs time for one fly.   \label{condmean}}
\end{figure}

Perhaps the most important prediction of our model is the probability of the behavioral state at time $t$ given the states at all other times $t'$,
\begin{eqnarray}
P\left(\sigma_t | \{\sigma_{t'\neq t}\}\right) &=& {{\exp\left[ h_{\rm eff}\left( \{\sigma_{t'\neq t}\}\right)\sigma_t\right]}\over{2\cosh[h_{\rm eff}\left( \{\sigma_{t'\neq t}\}\right)]}}\label{condheff}\\
h_{\rm eff}\left( \{\sigma_{t'\neq t}\}\right) &=& h + \sum_{t'\neq t} J(t-t')\sigma_{t'} .
\label{heff}
\end{eqnarray}
We can test this prediction by walking through the data and collecting all the moments in time where $h_{\rm eff}$ falls into some small range, and averaging the behavioral states over these times \cite{testh}; we should find $\langle\sigma\rangle= \tanh (h_{\rm eff})$.  To avoid any dangers of over--fitting \cite{overfit}, we assume that the system is described {\em exactly} by the inverse square Ising model, $J(\tau ) =J_0 /\tau^2$, and fit the one parameter $J_0$.  Results are shown  in Fig \ref{condmean}a.

The agreement between theory and experiment that we see in Fig \ref{condmean}a is very good, and  the best fit value of $J_0$ is consistent with the estimate of $J(\tau = 1)$ in Fig \ref{J}.  The largest deviations between theory and experiment are at extreme values of $h_{\rm eff}$, where behavior is even more nearly deterministic ($\langle\sigma\rangle \rightarrow \pm 1$) than predicted by the theory.  The distribution of $h_{\rm eff}$ is strongly bimodal, as if the fly were alternating between strong internal biases, and we see this in the time course of $h_{\rm eff}$  in Fig \ref{condmean}b.

Equation \ref{condheff} has the form of a behavioral state responding at each moment in time to a bias $h_{\rm eff}$, but these biases are determined by behavioral states at other times through Eq (\ref{heff}).   Importantly,  the full model in Eq (\ref{Pmaxent})  can be rewritten in a form where behavioral states respond independently at each moment in time to a completely internal, fluctuating  bias $\phi_t$ \cite{bialek_20},
\begin{equation}
P\left( \{\sigma_t\}\right) = \int {\cal D}\phi \, P\left( \{\phi_t\}\right) \prod_t {{\exp\left[\sigma_t ( h + \phi_t)\right]}\over{2\cosh( h + \phi_t)}} .
\end{equation}
The probability distribution of biases is given by
\begin{eqnarray}
P\left( \{\phi_t\}\right) &=& {1\over Z}\exp\left[ - S\left( \{\phi_t\}\right)\right]\\
S\left( \{\phi_t\}\right) &=& {1\over 2}\sum_{tt'} \phi_t K(t-t') \phi_{t'} - \sum_t \ln\cosh(h + \phi_t) ,\nonumber\\
&&
\end{eqnarray}
and the kernel $K(\tau)$ is the inverse of the interaction $J(\tau)$,
\begin{equation}
\sum_{t''} K(t-t'')J(t''-t') = \delta_{tt'} .
\end{equation}
For models in which $J(\tau ) \sim 1/\tau^2$, for large $\tau$ we have $K(\tau ) \sim -1/\tau^2$, so that the internal biases also must have interactions that extend over long spans of time;   there must be yet more internal degrees of freedom which carry memory across these spans.  To generate the $\sim 1/\tau^2$ behavior exactly requires that the dimensionality of these internal dynamics be effectively infinite \cite{bialek_20}.  In a different language, if we wanted to generate $\sim 1/\tau^2$ behavior exactly using a hidden Markov model, the number of hidden states would have to be effectively infinite as well.  While not suggesting a specific mechanism, the explicit $\sim1/\tau^2$ interactions provide a simple description of the phenomenology, the minimum required to match the observed correlations $C(\tau)$.

To summarize, we have tamed some of the complexity of the behavior in walking flies by compressing the behavioral states into binary variables which preserve correlations from one state to the next.  The resulting binary sequence nonetheless captures long ranged correlations in the behavior, with memory detectable out to thousands of transitions.  The simplest model that describes this behavior turns out to be an Ising model with almost perfectly inverse square interactions, and this model gives excellent quantitative predictions for the behavioral state at one moment in time given the surrounding sequence of states.  Although very simple, this description points to an effectively infinite number of hidden states.

\begin{acknowledgments}
This work was supported in part by the National Science Foundation through the Center for the Physics of Biological Function (PHY--1734030), the Center for the Science of Information (CCF--0939370), the NSF--Simons Center for Quantitative Biology (DMS--1764421), and Grant  PHY--1607612; by the National Institutes of Health (NS104889);  and by the Simons Foundation. 
\end{acknowledgments}


\begin{thebibliography}{99}
%
\bibitem{frisch_74}
K von Frisch, Decoding the language of the bee. {\em Science} {\bf 185,} 663--668 (1974).
%
\bibitem{gould_82}
JL Gould, {\em Ethology:  The Mechanisms and Evolution of Behavior} (WW Norton, New York, 1982).
%
\bibitem{lawson+uhlenbeck_50}
JL Lawson and GE Uhlenbeck, {\em Threshold Signals.}  MIT Radiation Laboratory Series 24.  (McGraw--Hill, New York, 1950).
%
\bibitem{green+swets_66}
DM Green and JA Swets, {\em Signal Detection Theory and Psychophysics} (Wiley,
New York, 1966).
%
\bibitem{osborne+al_05}
 LC Osborne, SG Lisberger, and W Bialek,   A sensory source for motor variation.  {\em Nature} {\bf 437,} 412--416 (2005).
%
\bibitem{stephens+al_08}
GJ Stephens, B Johnson--Kerner, W Bialek, and WS Ryu,    Dimensionality and dynamics in the behavior of {\em C. elegans}.    {\em PLoS Comput Biol} {\bf 4,} e1000028 (2008)
%
 \bibitem{ahamed+al_20}
T Ahamed, AC Costa, and GJ Stephens, Capturing the continuous complexity of behaviour in {\em C elegans}. {\em Nat Phys} in press (2020).\\
 https://doi.org/10.1038/s41567-020-01036-8.
%
\bibitem{branson+al_09} 
K Branson, AA Robie, J Bender, P Perona, and MH Dickinson, High--throughput ethomics in large groups of {\em Drosophila}. {\em Nat Methods} {\bf 6,} 451--457 (2009).
%
\bibitem{berman+al_14}
GJ Berman,  DM Choi, W Bialek, and JW Shaevitz, Mapping the stereotyped behaviour of freely moving fruit flies.  {\em J R Soc Interface} {\bf 11,} 20146072 (2014). 
%
\bibitem{berman+al_16}
GJ Berman, W Bialek, and JW Shaevitz, Predictability and hierarchy in {\em Drosophila} behavior.   {\em Proc Natl Acad Sci (USA)} {\bf 113,}  11943--11948 (2016).
%
\bibitem{wiltschko+al_15}
AB Wiltschko, MJ Johnson, G Iurilli, RE Peterson, JM Katon, SL Pashkovski, VE Abraira, RP Adams, and SR Datta,  Mapping sub--second structure in mouse behavior.  {\em Neuron} {\bf 88,} 1--15 (2015).
%
\bibitem{marshall+al_20}
JD Marshall, DE Aldarondo, TW Dunn, WL Wang, GJ Berman, and BP \"Olveczky,
Continuous whole-body 3D kinematic recordings across the rodent behavioral repertoire. {\em Neuron} in press (2020).\\ https://doi.org/10.1016/j.neuron.2020.11.016.
%
\bibitem{mathis+al_18}
A Mathis, P Mamidanna, KM Cury, T Abe, VN Murthy, MW Mathis, and  M Bethge, DeepLabCut: Markerless pose estimation of user-defined body parts with deep learning.  {\em Nat Neurosci} {\bf 21,} 1281--1289 (2018).
%
\bibitem{pereira+al_19}
T Pereira, D Aldarondo, L Willmore, M Kislin, SS Wang, M Murthy, and JW Shaevitz, Fast animal pose estimation using deep neural networks. {\em Nat Methods} {\bf 16,} 117--125 (2019).
%
\bibitem{brown+bivort_17}
AEX Brown and B de Bivort, Ethology as a physical science. {\em Nat Phys} {\bf 14,} 653--657 (2017).
%
\bibitem{datta+al_19}
SR Datta, DJ Anderson, K Branson, P Perona, and A Leifer, Computational neuroethology: A call to action. {\em Neuron} {\bf 104,} 11--24 (2019).
%
\bibitem{pereira+al_20}
TD Pereira, JW Shaevitz, and M Murthy, Quantifying behavior to understand the brain.  {\em Nat Neurosci} {\bf 23,} 1537--1549 (2020).
%
\bibitem{mathis+mathis_20}
MW Mathis and A Mathis, Deep learning tools for the measurement of animal behavior in neuroscience.
{\em Current Opin Neuro} {\bf 60,} 1--11 (2020).
%
\bibitem{bialek_20}
W Bialek, What do we mean by the dimensionality of behavior?  arXiv:2008.09574 [q--bio.NC] (2020).
%
\bibitem{tishby+al_99}
N Tishby, FC Pereira, and W Bialek,  The information bottleneck method. In {\em Proceedings of the 37th Annual Allerton Conference on Communication, Control and Computing,} B Hajek and RS Sreenivas, eds,  pp 368--377 (University of Illinois, 1999); arXiv:physics/0004057 (2000).
%
\bibitem{difference}
This result is different from that in Ref \cite{berman+al_16}, which used the (asymmetric) information bottleneck to find compressed representations that preserve the maximum information about the  future state, $I(z_t; x_{t+\tau})$.  In  the asymmetric case the optimal binary representation corresponds almost perfectly with moving vs idle.
%
\bibitem{Pc_C}
In comparing Figs \ref{Pc} and \ref{ss}, note that for binary variables we have ${\tilde P}_c (\tau ) = C(\tau )/[1 + \langle\sigma_t\rangle^2]$.
%
\bibitem{jaynes_57}
ET Jaynes, Information theory and statistical mechanics. {\em Phys Rev} {\bf 106,} 620--630 (1957).
%
\bibitem{bialek_12}
W Bialek, {\em Biophysics: Searching for Principles.}   (Princeton University Press, Princeton NJ, 2012).
%
\bibitem{tkacik+al_14}
G Tka\v{c}ik, O Marre, D Amodei, E Schneidman, W Bialek, and MJ Berry II,  Searching for collective behavior in a large network of sensory neurons.  {\em PLoS Comput Biol} {\bf 10,} e1003408 (2014).
%
\bibitem{anderson_84}
PW Anderson, {\em Basic Notions of Condensed Matter Physics} (Benjamin/Cummings, Menlo Park CA, 1984).
%
\bibitem{testh}
For a parallel discussion of states in a network of neurons: L Meshulam, JL Gauthier, CD Brody, DW Tank, and W Bialek,  Collective behavior of place and non-place neurons in the hippocampal network.    {\em Neuron} {\bf 96,} 1178--1191 (2017).
%
\bibitem{overfit}
Although we have a large amount of data, the model we are considering has 1000 parameters if we try to match correlations out to $\tau_{\rm max}=10^3$.  Precisely because correlations have a  long tail, it is not clear how many effectively independent samples we really have.  Thus, to test our theoretical approach, we take the simple inverse square model seriously.
%
\end{thebibliography}
\end{document}